\documentclass{article}
\usepackage{amsmath}
\usepackage[dvips]{epsfig}
\usepackage{amssymb}
\usepackage{fullpage}

\def\##1{{\bf #1}}
\def\=#1{\underline{\underline #1}}

\def\.{\mbox{ \tiny{$^\bullet$} }}

\def\eps{\varepsilon}
\def\epso{{\eps}_o}

\def\ux{\#u_x}
\def\uy{\#u_y}
\def\uz{\#u_z}
\def\zz{\uz\uz}

\def\muo{\mu_o}

\def\les{\left[}
\def\ris{\right]}
\def\lec{\left\{}
\def\ric{\right\}}

\def\n{^{(n)}}
\def\m{^{(m)}}
\def\eff{^{eff}}

\begin{document}

\noindent {\bf {\Large Constraints on effective constitutive parameters of certain bianisotropic laminated composite materials }}\\

 \vspace{10mm}

\noindent {\sf AKHLESH LAKHTAKIA} \\
Nanoengineered Metamaterials  Group \\
Department of Engineering Science \& Mechanics\\
212 Earth \& Engineering Sciences Building\\
Pennsylvania State University\\ University Park, PA 16802--6812\\
USA\\


\begin{abstract}  When the electrically thin unit cell of a laminated composite material is made of two bianisotropic sheets whose
constitutive properties in the thickness direction are decoupled from
the constitutive properties in the interfacial planes,  the laminated composite material can be homogenized into a material not
all of whose constitutive parameters are independent of each other. This non-independence of
the constitutive dyadics of the constituent materials and the homogenized composite material
is captured by two simple constraints, which may not hold if even one of the two constituent materials has more
complicated constitutive properties than stated above.

 \end{abstract}

\section{Introduction}
The electromagnetic response properties of laminated composite materials have been 
and continue to be of technoscientific
importance (Herpin, 1947,  Abel\`{e}s, 1950; Lafait \emph{et al.}, 1990; Neelakanta, 1995; 
Lakhtakia, 1996). A laminated composite material is made by stacking together sheets of 
different materials one on top of another, in order to form a unidirectionally nonhomogeneous 
material with piecewise-uniform constitutive properties. When a monochromatic plane wave 
is incident on this stratified composite material, standing waves are set up in each sheet, and 
electromagnetic energy may also be absorbed therein. In addition, the phenomenons of 
reflection and transmission occur. Across any interface, the tangential components of the electric 
and the magnetic field phasors  are continuous. These tangential components are the essential 
ingredients of a two-point boundary value problem that can be adequately stated and solved 
using matrix calculus. 
Whereas  2$\times$2 matrixes suffice for isotropic dielectric--magnetic sheets, 4$\times$4 
matrixes are needed for sheets with more complicated electromagnetic properties 
(Teitler \& Henvis, 1970; Mrozowski, 1986; Lakhtakia, 1987). 

A homogenization problem emerges when electrically thin sheets are stacked periodically
and the unit cell is also electrically thin (Wiener, 1912; Rytov, 1956; Rumsey, 1964).
 Such a periodically nonhomogeneous
composite material 
whose unit cell comprises two or more  sheets
\textit{may} be considered equivalent to a homogeneous material that necessarily has 
direction-dependent constitutive properties. Analytically straightforward
techniques can be employed to predict the effective (i.e.,
post-homogenization) constitutive parameters of a laminated composite material (Reese \& Lakhtakia, 1991;
Ramakrishna \& Lakhtakia, 2009). The effective constitutive parameters depend on the constitutive parameters of the constituent
materials as well as on their volume fractions.

Are all effective constitutive parameters necessarily independent of each other? Or, knowledge of some effective constitutive parameters
is sufficient to determine the others without recourse to the volume fractions of the constituent materials. This brief communication 
addresses this issue.

An $\exp(-i\omega t)$ time-dependence is implicit, with $\omega$ denoting the 
angular frequency. The permeability and permittivity of free space 
(i.e., vacuum) are denoted by $\muo$ and $\epso$, respectively. Vectors are in boldface, 
dyadics are   underlined twice, column vectors are in boldface and enclosed within square 
brackets, and matrixes are underlined twice and similarly bracketed. Cartesian unit vectors are 
identified as $\ux$, $\uy$ and $\uz$,
with the $z$ axis oriented normal
to the sheets. 

\section{Analysis}
Let us consider a periodic laminated composite material whose unit cell is made of two sheets of 
dissimilar, homogeneous, bianisotropic materials
labeled $1$ and $2$. The Tellegen constitutive
relations of the two materials are written in the frequency domain as (Lakhtakia, 1987)
\begin{equation}
\left.\begin{array}{l}
\#D=\epso\left(\=\eps\n\cdot\#E + \=\alpha\n\cdot\#H\right)\\[5pt]
\#B=\muo\left(\=\mu\n\cdot\#H +\=\beta\n\cdot\#E\right)
\end{array}
\right\}\,,\quad n\in\lec 1,2\ric\,.
\label{conrel-n}
\end{equation}
The relative permittivity dyadic and the relative permeability
dyadic of the $n$-th material are denoted by $\=\eps\n$ and $\=\mu\n$, respectively,
whereas the dyadics $\=\alpha\n$ and $\=\beta\n$ denote the magnetoelectric properties.
 The thickness of the  sheets made of the $n$-th constituent material is
denoted by $d_n$, so that
\begin{equation}
f_n= d_n/(d_1+d_2)\,,\quad n\in\lec 1,2\ric\,,
\end{equation}
is the volume fraction of the $n$-th material, with $f_2=1-f_1$.

Wave propagation in the laminated composite material can be handled by using the spatial Fourier transform as (Krowne, 1984)
\begin{equation}
\label{Fourier}
\left.\begin{array}{l}
\#E(x,y,z)=\int_{-\infty}^\infty\,\int_0^{2\pi}\,
\#e(z,\kappa,\psi) \exp\left[i \kappa (x\cos\psi+y\sin\psi)\right]\,d\psi\, d\kappa\\[5pt]
\#H(x,y,z)=\int_{-\infty}^\infty\,\int_0^{2\pi}\,
\#h(z,\kappa,\psi) \exp\left[i \kappa (x\cos\psi+y\sin\psi)\right]\,d\psi\,d\kappa
\end{array}
\right\}\,, 
\end{equation}
where  $\kappa$ is the spatial frequency
and $\psi$ is an angle. Substitution of this representation along with the
constitutive relations (\ref{conrel-n}) in the two Maxwell curl equations leads to the 4$\times$4-matrix
ordinary differential equation (Lakhtakia, 1987)
\begin{equation}
\frac{d}{dz}\les\#F(z,\kappa,\psi)\ris=i\les\=P^{(n)}(\kappa,\psi)\ris\.\les\#F(z,\kappa,\psi)\ris\,,
\quad n\in\lec1,2\ric\,,
\label{MODE}
\end{equation}
in any sheet made of the $n$-th material.  
The column vector
\begin{equation}
\les\#F(z,\kappa,\psi)\ris= \les\begin{array}{c} e_x(z,\kappa,\psi)\\ e_y(z,\kappa,\psi)\\ h_x(z,\kappa,\psi) \\ h_y(z,\kappa,\psi)\end{array}\ris
\end{equation}
represents components of the electric and magnetic field phasors that are tangential to the bimaterial
interfaces in the laminated composite material. The 4$\times$4 matrix $\les\=P\n(\kappa,\psi)\ris$ is
too cumbersome to reproduce here, but it can be put in the following form (Lakhtakia \& Weiglhofer, 1997):
\begin{equation}
\les\=P\n(\kappa,\psi)\ris=\les\=Q_1\n\ris +
\kappa\lec\les\=Q_2\n\ris\, e^{-i\psi} +
\les\=Q_3\n\ris \,e^{i\psi}\ric 
+
\kappa^2\les\=Q_4\n(\psi)\ris\,,
\quad n\in\lec 1,2\ric\,.
\end{equation}

Provided the sheets are electrically thin (Lakhtakia \& Krowne, 2003; Mackay, 2008), the laminated composite material
is equivalent to a homogeneous material with the following constitutive relations:
\begin{equation}
\left.\begin{array}{l}
\#D=\epso\left(\=\eps\eff\cdot\#E + \=\alpha\eff\cdot\#H\right)\\[5pt]
\#B=\muo\left(\=\mu\eff\cdot\#H +\=\beta\eff\cdot\#E\right)
\end{array}
\right\}\,.
\label{conrel-eff}
\end{equation}
Equations (\ref{Fourier}) and (\ref{MODE}) govern wave propagation in the homogenized composite material, except that the
matrix $\les\=P^{(n)}(\kappa,\psi)\ris$ must be replaced by
\begin{equation}
\les\=P\eff(\kappa,\psi)\ris=\les\=Q_1\eff\ris +
\kappa\lec\les\=Q_2\eff\ris\, e^{-i\psi}  +
\les\=Q_3\eff\ris\, e^{i\psi} \ric 
+
\kappa^2\les\=Q_4\eff(\psi)\ris\,.
\end{equation}

As the sheets and the unit cell are electrically thin, 
invocation of the long--wavelength approximation (Reese \& Lakhtakia, 1991;  Lakhtakia \& Krowne, 2003) yields
the relation
\begin{equation}
\les\=P\eff(\kappa,\psi)\ris = f_1\,\les\=P^{(1)}(\kappa,\psi)\ris + f_2\,\les\=P^{(2)}(\kappa,\psi)\ris\,,\quad
\forall\lec\kappa,\,\psi\ric\,,
\end{equation}
which can be used to determine $\lec\=\eps\eff,\,\=\mu\eff,\,\=\alpha\eff,\,\=\beta\eff\ric$ in terms
of $\lec\=\eps^{(1)},\,\=\mu^{(1)},\,\=\alpha^{(1)},\,\=\beta^{(1)}\ric$,
$\lec\=\eps^{(2)},\,\=\mu^{(2)},\,\=\alpha^{(2)},\,\=\beta^{(2)}\ric$,
$f_1$, and $f_2$.  The equation
\begin{equation}
\label{Q4}
\les\=Q_4\eff(\psi)\ris = f_1\,\les\=Q_4^{(1)}(\psi)\ris + f_2\,\les\=Q_4^{(2)}(\psi)\ris 
\end{equation}
is satisfied $\forall\psi\in\left[0,2\pi\right)$ by
\begin{equation}
\label{zz}
\left.\begin{array}{l}
\eps_{zz}\eff=\gamma \left(f_1\,\eps_{zz}^{(1)}\Phi^{(2,2)} + f_2\,\eps_{zz}^{(2)}\Phi^{(1,1)}\right)\\
\mu_{zz}\eff=\gamma \left(f_1\,\mu_{zz}^{(1)}\Phi^{(2,2)} + f_2\,\mu_{zz}^{(2)}\Phi^{(1,1)}\right)\\
\alpha_{zz}\eff=\gamma \left(f_1\,\alpha_{zz}^{(1)}\Phi^{(2,2)} + f_2\,\alpha_{zz}^{(2)}\Phi^{(1,1)}\right)\\
\beta_{zz}\eff=\gamma \left(f_1\,\beta_{zz}^{(1)}\Phi^{(2,2)} + f_2\,\beta_{zz}^{(2)}\Phi^{(1,1)}\right)
\end{array}\right\}\,,
\end{equation}
where $\eps_{zz}\eff=\uz\cdot\=\eps\eff\cdot\uz$, etc., and
\begin{equation}
\left.\begin{array}{l}
\Phi^{(n,m)}=\eps_{zz}\n\,\mu_{zz}\m-\alpha_{zz}\n\,\beta_{zz}\m\\
\gamma^{-1}=f_1^2\,\Phi^{(2,2)}+f_2^2\,\Phi^{(1,1)}
+f_1f_2\left(\Phi^{(1,2)}+\Phi^{(2,1)}\right)
\end{array}\right\}\,.
\end{equation}
 Solution of the equations
\begin{equation}
\label{Q1Q2Q3}
\les\=Q_m\eff\ris = f_1\,\les\=Q_m^{(1)}\ris + f_2\,\les\=Q_m^{(2)}\ris\,,\quad
m\in\lec1,2,3\ric,
\end{equation}
yields analytical expressions for the remaining components
of $\lec\=\eps\eff,\,\=\mu\eff,\,\=\alpha\eff,\,\=\beta\eff\ric$, but  those expressions are far too unwieldy, in general, for
reproduction here.

\subsection{A special case}\label{special}
A special case emerges when all four constitutive dyadics of both constituent materials are of the
form
\begin{equation}
\left.\begin{array}{l}
\=a\n=a_{zz}\n\zz + \=A\n\\
 \uz\cdot\=A\n=\=A\n\cdot\uz=\#0
 \end{array}\right\}\,,\quad a\in\lec\eps,\mu,\alpha,\beta\ric\,,\quad n\in\lec 1,2\ric,
\end{equation}
so that $a_{zx}\n=a_{zy}\n=a_{xz}\n=a_{yz}\n= 0$. Then, $\les\=Q_2\n\ris=\les\=Q_3\n\ris=\les\=0\ris$,
$(n\in\lec1,2\ric)$, and
the simplifications
\begin{equation}
a_{zx}\eff=a_{zy}\eff=a_{xz}\eff=a_{yz}\eff= 0\,,\quad
a\in\lec\eps,\mu,\alpha,\beta\ric\,,
\end{equation}
follow from eqs.~(\ref{Q1Q2Q3}). Furthermore, those equations yield
\begin{equation}
\label{pq}
a_{pq}\eff=f_1\,a_{pq}^{(1)}+f_2\,a_{pq}^{(2)}\,,
\quad a\in\lec\eps,\mu,\alpha,\beta\ric\,,
\quad p\in\lec x,y\ric\,,
\quad q\in\lec x,y\ric\,.
\end{equation}
These expressions hold in addition to eqs. (\ref{zz}) for $a_{zz}\eff$.

Equations (\ref{zz}) and (\ref{pq}) indicate that---once $\eps_{pq}\eff$
and $\mu_{pq}\eff$, $\left( p\in\lec x,y,z\ric,\, q\in\lec x,y,z\ric\right)$,
have been calculated by using the constitutive parameters and the volume fractions
of both constituent materials---$\alpha_{pq}\eff$
and $\beta_{pq}\eff$ can be obtained without using the volume fractions at all. Thus,
10 of the 20 effective constitutive parameters in the special case under
consideration require knowledge of the volume fractions, but thereafter the remaining 10
effective constitutive parameters do not.  Accordingly, all effective constitutive parameters are not independent of each other.

\subsection{General consideration}\label{general}

Equations (\ref{zz}) and (\ref{pq}) suggest the formulation of the constraints
\begin{equation}
\label{rel}
{\rm Det}
\left[\begin{array}{ccc}
\=\eps^{(1)}   &\=\eps^{(2)} & \=\eps\eff\\[5pt]
\=\mu^{(1)}   &\=\mu^{(2)} & \=\mu\eff\\[5pt]
\=\alpha^{(1)}   &\=\alpha^{(2)} & \=\alpha\eff
\end{array}\right]=
{\rm Det}
\left[\begin{array}{ccc}
\=\eps^{(1)}   &\=\eps^{(2)} & \=\eps\eff\\[5pt]
\=\mu^{(1)}   &\=\mu^{(2)} & \=\mu\eff\\[5pt]
\=\beta^{(1)}   &\=\beta^{(2)} & \=\beta\eff
\end{array}\right]=
0\,.
\end{equation}
Both certainly hold true
for the special case treated in  Sec.~\ref{special}. Could these constraints hold in a more general sense?

When even one of the two constituent materials is more general than in Sec.~\ref{special}, analytical
expressions for $\lec\=\eps\eff,\,\=\mu\eff,\,\=\alpha\eff,\,\=\beta\eff\ric$ turn out to be so  huge
that analytical expressions of the determinants in the two
constraints (\ref{rel}) could not be manipulated to ascertain if both constraints hold true in general.

Several numerical experiments were conducted,
wherein all 36 effective constitutive parameters  were computed with $f_1\in\left(0,1\right)$ and with
 none of the 72 constitutive parameters of the two constituent materials
of null value. The determinants  in eqs.~(\ref{rel}) turned out be significantly different from zero,
indicating thereby that the constraints (\ref{rel}) do not hold in general.

\subsection{Unit cell with 3 or more sheets}
Suppose the unit cell of a certain laminated composite material comprises three different sheets.
Two adjacent sheets can be homogenized into one thicker sheet, and the conclusions in Sec.~\ref{special}
and \ref{general} apply to that homogenization. This thicker sheet and the third sheet can also
be homogenized into a single sheet of the thickness of the unit cell, and the conclusions in Sec.~\ref{special}
and \ref{general} also apply to that homogenization. This procedure can be adopted  for unit cells with 4 or
more sheets, so long as the unit cell, after all homogenization steps, remains electrically thin.

\section{Concluding Remarks}
Suppose that the unit cell of a laminated composite material is made of two bianisotropic sheets whose
constitutive properties in the thickness direction are decoupled from
the constitutive properties in the interfacial planes. Provided that the long-wavelength
approximation is applicable, the laminated composite material can be homogenized into a material
all of whose constitutive parameters are not independent of each other. This non-independence of
the constitutive dyadics of the constituent materials and the homogenized composite material
is captured by two simple constraints. When even one of the two constituent materials has more
complicated constitutive properties, the constraints are not expected to hold. 

Given that magnetoelectric properties are considerably rarer and usually weaker than dielectric-magnetic
properties, the two constraints, when valid, should
be considered as applicable on the magnetoelectric dyadics. Isotropic chiral and biisotropic materials,
general uniaxial bianisotropic materials,
gyrotropic materials such as ferrites and Faraday chiral materials, etc., (Mackay \&
Lakhtakia, 2008) all lie within the scope of the two
constraints.
In the realm of classical electromagnetics (i.e., at frequencies not exceeding about 750~THz), we can
expect the homogenization procedure to hold for sheets as thin as 10 molecular diameters (Kim {\em et al.}, 2005).
Thus, the conclusions drawn in this communication can also be expected to hold at optical and lower frequencies for
many types of nanotextured thin films.

\vspace{5 mm}
\noindent {\bf Acknowledgments.} Thanks are due to S. Anantha Ramakrishna (IISER, Mohali, India) for
discussions. The Charles Godfrey Binder Endowment at Penn State is gratefully acknowledged for partial
financial support.

\vspace{10mm}

 \noindent{\bf References}\\

\noindent Abel\`{e}s, F. 1950. Recherches sur la propagation des ondes \'{e}lectromagn\'etiques sinuso•dales dans les milieux 
stratifi\'{e}s. Application aux couches minces (1re partie). \emph{Ann. Phys. (Paris)} 5:596--640.\footnote{See Lakhtakia (1996)
for a facsimile reproduction.}\\

\noindent Herpin, A.  1947. Calcul du pouvoir r\'{e}flecteur d'un syst\`{e}me stratifi\'{e} quelconque. \emph{C. R. Acad. Sci. (Paris)}
225:182--183.\\

\noindent Kim, H.-Y., J.O. Sofo, D. Velegol, M.W. Cole, and G.~Mukhopadhyay. 2005.
Static polarizabilities of dielectric nanoclusters. \emph{Phys. Rev. A} 72:053201.\\

\noindent Krowne, C.M. 1984. Fourier transformed matrix method of finding
propagation characteristics of complex anisotropic
layered media. \emph{IEEE Trans. Microw. Theory Tech.} 32:1617--1625.\\

\noindent
Lafait, J., T. Yamaguchi, J. M. Frigerio, A. Bichri, and K. Driss-Khodja. 1990.
Effective medium equivalent to a symmetric multilayer at oblique incidence. \emph{Appl. Opt.} 29:2460--2464.\\

\noindent Lakhtakia, A. 1987. Cartesian solutions of Maxwell's equations for linear, anisotropic media---Extension
of the Mrozowski algorithm. \emph{Arch. Elektr. \"Uber.} 41:178--179.\\

\noindent Lakhtakia, A. (ed.) 1996. \emph{Selected papers on linear optical composite materials}. Bellingham, WA, USA: SPIE.\\

\noindent Lakhtakia, A., and C.M. Krowne. 2003.
Restricted equivalence of paired epsilon-negative
and mu-negative layers to a negative phaseÐvelocity
material (\textit{alias} left-handed material). \emph{Optik} 114:305--307.\\

\noindent Lakhtakia, A., and W.S. Weiglhofer. 1997.
Green function for radiation and propagation in
helicoidal bianisotropic mediums. \emph{IEE Proc.--Microw. Antennas Propagat.}
144:57--59.\\

\noindent Mackay, T.G. 2008. Lewin's homogenization formula revisited for
nanocomposite materials. \emph{J. Nanophoton.} 2:029503.\\

\noindent Mackay, T.G., and A. Lakhtakia. 2008. Electromagnetic fields in linear
bianisotropic mediums. \emph{Prog. Opt.} 51:121--209.\\

\noindent Mrozowski, M. 1986. General solutions to Maxwell's equation in a bianisotropic medium.---A
computer oriented, spectral domain approach. \emph{Arch. Elektr. \"Uber.} 40:195--197.\\

\noindent Neelakanta, P.S. 1995.
\emph{Handbook of composite materials}.
Boca Raton, FL, USA: CRC Press.\\

\noindent Ramakrishna, S.A., and A. Lakhtakia. 2009.
Spectral shifts in the properties of a
periodic multilayered stack due to
isotropic chiral layers. \emph{J. Opt. A: Pure Appl. Opt.} (accepted for publication).\\

\noindent Reese, P.S., and A. Lakhtakia. 1991. Low-frequency electromagnetic
properties of an alternating stack of thin uniaxial dielectric
laminae and uniaxial magnetic laminae. \emph{Z. Nat\"urforsch.
A} 46:384--388.\\

\noindent Rumsey, V.H. 1964. Propagation in generalized gyrotropic media.
\emph{IEEE Trans. Antennas Propagat.} 12:83--88\\

\noindent Rytov, S.M. 1956. Electromagnetic properties
of a finely stratified medium. \emph{Sov. Phys. JETP} 2:466--475.\\

\noindent Teitler, S., and B.W. Henvis. 1970. Refraction in stratified, anisotropic media. \emph{J. Opt. Soc. Am.}
60:830--834.\\

\noindent Wiener, O. 1912. Die Theorie des Mischk\"orpers f\"ur das Feld der Station\"aren Str\"omung.
Erste Abhandlung: Die Mittelwerts\"atze f\"ur Kraft, Polarisation und Energie. \emph{Abh.
 Math.-Phys. Kl. K\"onigl. SŠchs. Ges. Wissen.} 32:507--604.\footnote{See Lakhtakia (1996) for Bernhard
 Michel's synposis in English of this landmark paper.}\\

\end{document}